\documentclass[12pt,preprint]{aastex}
\usepackage{psfig}
\newcommand{\beq}{\begin{equation}}
\newcommand{\eeq}{\end{equation}}
\newcommand{\beqn}{\begin{eqnarray}}
\newcommand{\eeqn}{\end{eqnarray}}

\newcommand{\no}{\noindent}

\begin{document}

\title{\bf{On the Mass-Period Distributions and
Correlations of Extrasolar Planets}}

\author{Ing-Guey Jiang$^a$, Li-Chin Yeh$^b$, Yen-Chang Chang$^b$,
 Wen-Liang Hung$^c$}

\affil{{\small $^a$Department of Physics, National Tsing Hua University, Hsin-Chu, Taiwan}\\
{\small $^b$Department of Applied Mathematics,
National Hsinchu University of Education, Hsin-Chu, Taiwan} \\
{\small $^c$Graduate Institute of Computer Science, National
Hsinchu University of Education, Hsin-Chu, Taiwan}}

\begin{abstract}
In addition to fitting the data of 233 extra-solar planets with power
laws,
we construct a correlated
mass-period distribution function of extrasolar planets,
as the first time in this field. The algorithm to generate 
a pair of positively correlated beta-distributed random variables
is introduced and used for the construction of 
correlated distribution functions.
We investigate the mass-period correlations of extrasolar
planets both in the linear and logarithm spaces,
determine the confidence intervals of the correlation coefficients,
and confirm that there is a
positive mass-period correlation for the extrasolar
planets. In addition to the paucity of massive close-in planets, which
makes the main contribution on this correlation, there are other fine
structures for the data in the mass-period plane.
\end{abstract}


\section{Introduction}

The observational effort has led to the discovery of
more than 200 extrasolar planets (exoplanets), in the mass range
0.03 to 20 Jupiter Mass ($M_J$), with orbital periods from a few days
to about 4000 days. Many interesting problems about the formation
and evolution of planetary systems have therefore been studied with
the information provided by these detected systems
(Jiang \& Ip 2001,
Laughlin \& Chambers 2001, Kinoshita \& Nakai 2001,
Gozdziewski \& Maciejewski 2001, Ji et al. 2002, Ji et al. 2003,
Jiang \& Yeh 2004)

Moreover, due to the growing number of detected extrasolar planets, several
groups have been working on the statistical distributions and
possible correlations. Assuming that the mass and period
distributions are two independent power-law functions, Tabachnik
and Tremaine (2002) used the method of maximum likelihood to
determine the best power-index. Though they found that
the uncertainties in the mass and period
distributions are coupled, the study of the possible mass-period
correlations is beyond their scope due to the principal assumption
of two independent power-law functions.

In addition, Zucker and Mazeh (2002) calculated
the correlation
coefficient between mass and period for the detected data in the
${\rm ln}P-{\rm ln}M$ space. They used Monte Carlo simulation
to determine the $p$-value for testing whether the correlation is
significant or not. They concluded that the mass-period
correlation is significant.
However, at the time of their study, the number of detected exoplanets
was limited, so only 66 planets were used. 

Over the years, much more exoplanets with different properties 
are discovered. For example, more hot Jupiters are found 
due to the effort of transit
surveys, a few newly detected exoplanets are 
moving on extremely eccentric orbits,
and the exoplanets with mass in the order of Earth Mass are also discovered.
These results have in fact brought this field into a completely new era.
Jiang et al. (2006) did cluster analysis on 143 samples and found that 
the data grouping could be related with the dynamical processes of planetary
systems.
This approach was agreed by
Marchi (2007), in which an extrasolar planet taxonomy
was presented.

Therefore, it is about the time to construct a new distribution function. 
As in Tabachnik and Tremaine (2002), assuming 
the mass and period are 
two independent power-law distributions, we use 233 samples of exoplanets
from exoplanet website (http://exoplanet.eu/catalog-all.php) 
on 6th July 2007 to construct updated 
distribution functions. Our samples do not include
OGLE235-MOA53 b,
2M1207b,
GQLup b,
HD 187123 c,
ABPic b,
SCR 1845 b,
SWEEPS-04, due to the missing of either their periods or
mass. The three 
outliers,
PSR 1257+12 b, 
HD 154345 b, 
PSR B1620-26 b with either extremely small mass or huge periods
are also excluded. 
(Please note that the data of mass 
means the value of projected mass in this paper.) 

Moreover, because the mass and period are likely to be correlated
according to the results of Zucker and Mazeh (2002), a new distribution
function without the assumption that the mass and period are
independent would be more satisfactory.
That is, we hope to construct a new distribution function, in which
the mass and period can be coupled. This was not possible until 
an algorithm for generating two positively
correlated beta-distributed random variables was provided in
Magnussen (2004). We therefore have to employ the beta distribution for this
part of calculations.

After that, in order to have a careful investigation on the possible 
mass-period
correlations, we work on the correlation coefficients for the data 
both in the linear and logarithm spaces. 
A standard
method in statistics called the bootstrap method will
be used to get the confidence
intervals of correlation coefficients. 

In the
following, we present the construction a new mass-period
power-law distribution function in Section 2
and the correlated mass-period distribution function
in Section 3. We present the bootstrap method in Section 4
and describes the results of the correlation coefficients and
confidence intervals in Section 5.
Finally, we provide
the concluding remarks in the final section.

\section{The Power-Law Distribution Function}

In this
section, we construct a new distribution function, assuming that the mass
and period are two independent power laws. 
We consider the probability density function (pdf) of
the power law has the following form
\begin{eqnarray}
f_{power}(x|k)=C(k) x^{-k},\,\, 0<a< x< b<\infty,
\end{eqnarray}
where the exponent $k$ is an unknown parameter and the constant
$C(k)$ is given by the normalization requirement that
$$1=\int_a^b f_{power}(x|k)dx=\frac{C(k)}{1-k}(b^{1-k}-a^{1-k}).$$
That is,
$$C(k)=\frac{1-k}{b^{1-k}-a^{1-k}}.$$

When sampling is from a population described by (1), 
the parameter $k$ yields the knowledge of the entire population.
Hence, it is natural to seek a method of finding a good estimator
of $k$. The method of maximum likelihood is one of the most
popular techniques for deriving estimators. Let $X_1,\cdots, X_n$
be independent and identically distributed (i.i.d.) samples from
the pdf $f_{power}(x|k)$, the likelihood function is given by
$$L(k|x_1,\cdots,x_n)=\prod_{i=1}^n f_{power}(x_i|k).$$
Taking logarithm of both sides, differentiating partially with
respect to the parameter $k$, and setting the result to zero we
can determine the maximum likelihood estimate (MLE) of $k$ by
solving
\begin{eqnarray}
\frac{a^{1-k}-b^{1-k}+(1-k)(b^{1-k}\ln b-a^{1-k}\ln
a)}{(1-k)(b^{1-k}-a^{1-k})}=\frac{\sum_{i=1}^n \ln x_i}{n}.
\end{eqnarray}

Now, we use the power-law distributions
$$f^M_{power}(m|k_m)=C(k_m) m^{-k_m},\,a_m<m<b_m$$
and
$$f^P_{power}(p|k_p)=C(k_p) p^{-k_p},\,a_p<p<b_p$$
 to fit the 233 observed
data in the $M$ and $P$ spaces, respectively.
First, we choose the range of $M$
as follows:
$$a_m=M_{min},\,\,b_m=M_{max},$$
where $M_{min}$ and $M_{max}$ are the smallest and largest mass
of the data set. That is, $a_m=0.012$ and $b_m=18.4$. By Eq. (2), we
obtain the MLE of $k_m$, which is $\hat k_m=0.7805$. Similarly, 
the range of $P$ is $a_p=1.211909$, $b_p=4517.4$ and the
MLE of $k_p$ is $\hat k_p=0.9277$.

The histograms of
observed data in the $M$ and $P$ spaces are showed in  
Fig. 1. Fig.1(a) is the one for $M$ and the area covered by this 
 histogram is 115.5. We define the curve,
$$f^{his_M}_{power}\equiv 115.5 f^M_{power}(m|\hat k_m)$$
and plot it as the dotted curve in  Fig.1(a) for comparison.
 Similarly, the histogram in Fig. 1(b) is for $P$ and its area is 11650.
We define another curve,
$$f^{his_P}_{power}\equiv 11650 f^P_{power}(p|\hat k_p)$$
and plot it as the dotted curve in  Fig.1(b).

\section{The Correlated Distribution Function}


In \S 2, the distributions of mass and period 
of the extrasolar planets are described by two independent power laws.
However, using a data set of 66 exoplanets, Zucker and Mazeh (2002)
suggested the possible mass-period correlation.
Further,  
our data of 233 samples here show that the
correlation coefficient in $M-P$ space is $0.1762$ and
this indicates there exists a positive correlation
between $M$ and $P$. 
It is thus not suitable to use two
independent power laws to describe the joint
mass-period distribution.
Therefore, we need
to know how to simultaneously describe and use probability models
to elicit information from the mass and period measurements.
It is necessary to
construct a new distribution function, in which the mass and
period can be positively correlated and coupled. This was not
possible until an algorithm for generating two positively
correlated beta-distributed random variables was provided in
Magnussen (2004). This is the reason why we decide to proceed
with the beta distribution here.

The beta distributions are very general, have many possible
different functional shapes, and have the advantage that the
variable boundaries and the normalizations are automatically
considered. The beta distributions are continuous on the finite
interval $(c,d)$, $-\infty<c<d<\infty$, indexed by two positive
parameters $\alpha$, $\beta$. The pdf is given by
\begin{eqnarray}
f_{beta}(x|\alpha,\beta)=\displaystyle\frac{1}{B(\alpha,\beta)}
\displaystyle\frac{(x-c)^{\alpha-1}(d-x)^{\beta-1}}{(d-c)^{\alpha+\beta-1}},
\,\,c< x< d,\,\alpha>0,\,\beta>0,
\end{eqnarray}
where $B(\alpha,\beta)$ denotes the beta function,
$$B(\alpha,\beta)=\int_0^1 t^{\alpha-1}(1-t)^{\beta-1}dt.$$

In Eq.(3), the pdf $f_{beta}(x|\alpha,\beta)$ satisfies
 $\int_c^d f_{beta}(x|\alpha,\beta)dx=1$. The beta function $B(\alpha,\beta)$ can be
expressed as
$$B(\alpha,\beta)=\displaystyle\frac{\Gamma(\alpha)\Gamma(\beta)}{\Gamma(\alpha+\beta)},$$
where the gamma function is
\beq
  \Gamma(\alpha)=\int_0^{\infty} t^{\alpha-1} e^{-t}dt.{\label{eq:gamma}}
\eeq
Considering the following transformation
$$y=\displaystyle\frac{x-c}{d-c},$$
we then have the pdf
\begin{eqnarray}
f(y|\alpha,\beta)=\displaystyle\frac{1}{B(\alpha,\beta)}
 y^{\alpha-1}(1-y)^{\beta-1},\,\, 0< y< 1,
\end{eqnarray}
which is called the standard beta distribution.

The beta distribution is often used to model a phenomenon
which could be described by the values of random variables defined
in a finite interval. As the parameters $\alpha$ and $\beta$ vary,
the beta distribution takes on many shapes, as shown in Fig. 2.
The pdf can be strictly increasing, strictly decreasing, or
U-shaped. The case $\alpha=\beta$ yields a pdf symmetric about
$(c+d)/2$. If $\beta=1$, the distribution is called a
power-function distribution. That is, this distribution is one
kind of power-law functions.

From the above discussion, it is clear that the beta distributions
are very versatile and can be used for many different purposes.
This flexibility encourages its empirical use in a wide range of
applications. For example,  Wall et al. (2000) successfully used
the beta distribution to model both the subgrid-scale pdf and the
subgrid-scale Favor pdf of the mixture fraction. Ettoumi et al.
(2002) used beta distributions to analyze solar measurements in
Algeria. Flynn (2004) suggested the beta distribution as a
suitable model for human exposure to airborne contaminants. Ji et
al. (2005) proposed a beta-mixture model to analyze a large number
of correlation coefficients in bioinformatics.

The standard beta function is for one variable only. If there are
more than one variables and they are independent to each other,
the extension to multi-variable cases is straightforward. However,
to have a generalized beta function with a pair of correlated
variables is not easy. A few algorithms have been proposed to
generate pairs of correlated beta-distributed random variables
numerically (please see Johnson 1987, Loukas 1984, Michael \&
Schucany 2002). Due to the limitations of these algorithms, they
can only be used for particular types of data set. The algorithm
in Magnussen (2004) can generate a pair of positively correlated
beta-distributed random variables without any limitations. We thus
use it to construct the numerical mass-period distribution
function $f(M,P|\alpha_m,\beta_m,\alpha_p,\beta_p)$.

The probability that a planet with mass and orbital
period in the range $[M,M+dM]$, $[P,P+dP]$ is given by
$$f(M,P|\alpha_m,\beta_m,\alpha_p,\beta_p)dMdP,$$
where the marginal distributions of $M$ and $P$ follow the beta
distribution as in Eq.(3) with parameters $(\alpha_m, \beta_m)$
and $(\alpha_p, \beta_p)$, respectively. From the data, the
boundaries can be set such that $m_1< M< m_2$,
 $p_1< P< p_2$. Then, the marginal distributions of
the random variables
 $$M_1=\frac{M-m_1}{m_2-m_1}~~\mbox{and}~~P_1=\frac{P-p_1}{p_2-p_1}$$
 should follow the standard beta distributions as Eq.(5) with parameters
 $(\alpha_m, \beta_m)$ and $(\alpha_p, \beta_p)$, respectively.

Let us define that
\begin{eqnarray}
\delta_1=\rho(M_1,P_1)\times (1+\alpha_m+\alpha_p)\times C
 \end{eqnarray}
and
\begin{eqnarray}
\delta_2=\rho(M_1,P_1)\times (1+\beta_m+\beta_p)\times C,
 \end{eqnarray}
where
\begin{eqnarray}
C=\frac{\sqrt{\alpha_m\alpha_p\beta_m\beta_p(1+\alpha_m+\beta_m)(1+\alpha_p+\beta_p)}}
{(1+\alpha_p)(1+\beta_m)(1+\beta_p)+\alpha_m(1+\beta_m+\beta_p+\beta_m\beta_p+
\alpha_p(1+\beta_m+\beta_p))},
\end{eqnarray}
and $\rho(M_1, P_1)$ is the correlation coefficient
between $M_1$ and $P_1$.
Then, $M_1$ and $P_1$ generated by below equations would be a
pair of correlated beta-distributed variables:
\begin{eqnarray}
M_1=\frac{G(\alpha_m^*)+G(\delta_1)}{G(\alpha_m^*)+G(\delta_1)+
G(\beta_m^*)+G(\delta_2)}
\end{eqnarray}
and
\begin{eqnarray}
P_1=\frac{G(\alpha_p^*)+G(\delta_1)}{G(\alpha_p^*)+G(\delta_1)+
G(\beta_p^*)+G(\delta_2)},
\end{eqnarray}
where $G(\alpha)$ is a random variable distributed as a
gamma distribution with parameters $\alpha$ and $1$ and
$\alpha_m^*$, $\beta_m^*$, $\alpha_p^*$, $\beta_p^*$
are defined by
\beq
\alpha_m^* = \alpha_m - \delta_1
\eeq
\beq
\beta_m^* = \beta_m - \delta_2
\eeq
\beq
\alpha_p^* = \alpha_p - \delta_1
\eeq
\beq
\beta_p^* = \beta_p - \delta_2.
\eeq
Note that the pdf of a
gamma distribution with parameters $\alpha$ and $\beta$ is
(Hogg \& Craig 1989)
\beq
f(x)=\frac{1}{\Gamma(\alpha) \beta^{\alpha}} x^{\alpha-1} e^{-x/\beta},
\eeq
where $0 < x < \infty$.

The above procedure to generate a pair of positively correlated
beta-distributed $M_1$ and $P_1$ variables can be summarized as:

\noindent
{\bf Magnussen Algorithm}
\begin{itemize}
\item[Step 1] Assume that the marginal distribution of $M_1$,
i.e. $f_{M_1}(m|\alpha_m,\beta_m)$, is
a standard beta distribution with parameters
$(\alpha_m, \beta_m)$. Through the maximum likelihood method, we employ the
data to get the best estimation $(\hat\alpha_m,\hat\beta_m)$
of $(\alpha_m,\beta_m)$. Similarly, we also get the best estimation
$(\hat\alpha_p,\hat\beta_p)$ of $(\alpha_p,\beta_p)$ for $P_1$.

\item[Step 2] Calculate the value of $C$ by Eq.(8).
\item[Step 3]
Calculate the correlation coefficient $\hat\rho(M_1, P_1)$
from the data and use
it as the value of $\rho(M_1, P_1)$.

\item[Step 4] Calculate $\delta_1, \delta_2$ by Eqs.(6)-(7).

\item[Step 5] Calculate  $\alpha_m^*$, $\beta_m^*$,
$\alpha_p^*$, $\beta_p^*$ by Eqs.(11)-(14).

\item[Step 6] Generate pairs of $M_1$, $P_1$ by Eqs.(9)-(10).
\end{itemize}

We now apply the above algorithm on the data set of 233
exoplanets. To avoid the possible singularity,
the range
of $M$ is chosen to be $(M_{min}/1.5, 1.5M_{max})$.
That is, $m_1=0.008, m_2=26.7$. By the same reason, the range of
$P$ is taken as $p_1=P_{min}/1.5=0.8079, p_2=1.5P_{max}=6776.1$,
where $P_{min}$ and $P_{max}$ are the smallest and largest of
period of the observed data. Thus, the MLE of $\alpha_m$,
$\beta_m$, $\alpha_p$ and $\beta_p$ are $\hat\alpha_m=0.6524$,
$\hat\beta_m=5.9076$, $\hat\alpha_p=0.3697$, $\hat\beta_p=3.8445$,
respectively.  In addition, the data shows that mass-period
correlation coefficient $\hat\rho(M_1, P_1)=0.1762$. We then get
all the parameters' values as $C=0.2092$, $\delta_1=0.0745$,
$\delta_2=0.3961$, and $\alpha_m^*=0.5779$,$\beta_m^*=5.5115$,
$\alpha_p^*=0.2952$, $\beta_p^*=3.4484$.

Because the area of the histogram for $M$ in Fig. 1(a) is 
115.5, we define the curve,
$$f^{his_M}_{beta}\equiv 115.5 f_{beta}(m|\hat\alpha_m,\hat\beta_m)$$
and plot it as the solid curve in Fig.1(a) for comparison.
Similarly, the area of the histogram for $P$ in Fig. 1(b) 
is 11650, so
we define another curve,
$$f^{his_P}_{beta}\equiv 11650 f_{beta}(p|\hat\alpha_p,\hat\beta_p)$$
and plot it as the solid curve in  Fig.1(b).

These plots indicate that the beta
distribution presents a better fitting with the data, comparing
with the power law.
Due to that there is no closed form for the positively correlated
beta distribution, we  numerically  plot the
joint distribution of $M$ and $P$ as shown in Fig. 3.
Fig. 3 presents the three dimensional
plot of our correlated mass-period distribution function.
Fig. 4 shows the contour of it in
smaller ranges of mass and period.
We thus successfully construct, for the first time in this field, 
the correlated mass-period distribution function.

Please note that,
for pairs of quantities $(x_i, y_i), i=1,\ldots,n$, the
correlation coefficient $\hat\theta$ is usually given by \beq
 \hat\theta=\displaystyle\frac{\sum_{i=1}^n
  (x_i-\bar x)(y_i-\bar y)}
   {\sqrt{\sum_{i=1}^n (x_i-\bar x)^2 \sum_{i=1}^n (y_i-\bar y)^2}},
\label{eq:theta}
\eeq
where $\bar x=\sum_{i=1}^n x_i/n$,
 $\bar y=\sum_{i=1}^n y_i/n$. The value of $\hat\theta$ lies between $-1$
and $1$. If the data points completely lie on a straight line with
positive (negative) slope, the correlation coefficient
$\hat\theta=1$ ($\hat\theta=-1$) and it is called
 ``complete positive correlation'' (``complete negative correlation'').
When the data points are randomly distributed, the variables $x$
and $y$ are uncorrelated and $\hat\theta$ is near zero. Thus, the
value $\hat\theta$ is regarded as one conventional way to
quantitatively describe the strength of relationship between $x$
and $y$. Thus, our $\hat\rho(M_1, P_1)=0.1762$ was obtained by the
above equation.

\section{The Bootstrap Confidence Intervals}

To assess the statistical significance of the possible correlation, 
it would be good to have
the confidence interval corresponding to a given confidence level.
We use the bootstrap method to construct confidence intervals here.
Statistically, we determine the characteristics of the population
by taking samples. Since the sample represents the population,
analogous characteristics of the sample should give us information
about the population characteristics. The bootstrap method
proposed by Efron (1979) is a simple and straight-forward method
to calculate the approximated biases, standard deviations, and
confidence intervals, for example. It gives the population
characteristics by taking samples repeatedly from the original
data set.

The bootstrap method is used for i.i.d. data. DiCiccio \& Efron
(1996) found that the bootstrap confidence intervals are more {\it
accurate} than the classical normal approximation intervals. The
standard bootstrap method for confidence interval can be described
as follows. Given an observed $\mbox{i.i.d.}$ sample ${\bf z}
=\{z_1,\cdots,z_n\}$ from
 an unknown distribution function $F$,
we want to construct a confidence
 interval for an interesting parameter $\theta=\theta(F)$ based on ${\bf z}$.
Let $\hat{F}$ be the empirical distribution function which is
defined to be the discrete distribution that assigns the
probability $1/n$ on each value $z_i$, $i=1,\cdots,n$. The key
idea of the bootstrap method is a bootstrap sample, which is
defined to be a random sample of size $n$ drawn from $\hat{F}$,
say
 ${\bf z}^\ast=\{ z_1^\ast$ $,\cdots,z_n^\ast\}$. That is, the
 bootstrap data points $z_1^\ast,\cdots,z_n^\ast$ are a random
 sample of size $n$ drawn  from the data set
$\{z_1,\cdots,z_n\}$.

Let $\hat \theta=s({\bf z})$ be an estimate of $\theta$.
Corresponding to a bootstrap data set ${\bf z}^\ast$ is a
bootstrap replication of $\hat \theta$,
 $$\hat\theta^\ast=s({\bf z}^\ast).$$
 The quantity $s({\bf z}^\ast)$ is the result of applying the same
 function $s(\cdot)$ to ${\bf z}^\ast$ as was applied to
  ${\bf z}$. For example, if $s({\bf z})$ is the sample mean
  $\bar z=\sum_{i=1}^n z_i/n$ then $s({\bf z}^\ast)$ is the mean of
  the bootstrap data set, $\bar z^\ast=\sum_{i=1}^n z_i^\ast/n$.
The bootstrap algorithm, described next, is a data-based
simulation procedure to obtain a good approximation of the
confidence interval for $\theta$.

\noindent
  {\bf Bootstrap Algorithm}
\begin{itemize}
\item[Step 1] Draw a `` bootstrap sample''
 $z^\ast=(z_1^\ast,\cdots, z_n^\ast)$ according to $\hat F$.

\item[Step 2] Evaluate $\hat \theta^\ast=s({\bf z}^\ast)$, where
 $s({\bf z}^\ast)$ is the value of $s({\bf z})$ based on
${\bf z}^\ast$.

  Repeat the previous two steps a large number of times,
say $B$ times, to obtain $\hat \theta_1^\ast,\cdots,
 \hat\theta_B^\ast$.

\item[Step 3] Sort $\hat \theta_1^\ast,\cdots,
 \hat\theta_B^\ast$ to be the ordered list $\hat \theta_{(1)}^\ast,\cdots,
 \hat\theta_{(B)}^\ast$.

\item[Step 4] Let $\hat \theta_B^{\ast(\alpha)}$, $0<\alpha<0.5$,
be the ($100\alpha$)th empirical percentile of the
$\hat\theta_{(b)}^\ast$, $b=1,\cdots,B$, that is, the $(\alpha B)$th
value in the ordered list of
$\hat\theta_{(1)}^\ast,\cdots,\hat\theta_{(B)}^\ast$.  \\
 Likewise, let $\hat\theta_B^{\ast(1-\alpha)}$ be the
the $(1-\alpha) B$th
value in the ordered list of $\hat\theta_{(1)}^\ast,\cdots,
\hat\theta_{(B)}^\ast$.

Then, the approximate $1-2\alpha$ confidence interval
for $\theta$ is $(\hat\theta_B^{\ast(\alpha)},\,\hat
\theta_B^{\ast(1-\alpha)})$.
\end{itemize}
\noindent

Moreover, if $B\alpha$ is not an integer, the following procedure
can be used. Assuming $0<\alpha<0.5$, let $k=[(B+1)\alpha]$, i.e.
the largest integer $\le (B+1)\alpha$. Then we define the
$\hat\theta_B^{\ast(\alpha)}$ and $\hat \theta_B^{\ast(1-\alpha)}$
by the $k$th  and $(B+1-k)$th  values of $\hat\theta_{(b)}^\ast$,
$b=1,\cdots,B$.

In this paper, the bootstrap algorithm is used to construct
confidence intervals for correlation coefficients. Assuming that
$n$ objects are sampled from a population and two numerical
characteristics are measured on each of them, we end up with
bivariate random sample points $z_1=(x_1,y_1), \cdots,
z_n=(x_n,y_n)$. Let $\theta$ be the population correlation
coefficient. Based on  $n$ data points $z_1, \cdots, z_n$, the
population correlation coefficient $\theta$ is estimated by the
sample correlation coefficient $\hat\theta$ defined in
Eq.(\ref{eq:theta}). Independent repetitions of the bootstrap
sampling process give $B$ bootstrap replications
$\hat\theta_1^\ast,\cdots,\hat\theta_B^\ast$. Then we obtain the
 approximate $1-2\alpha$ confidence interval
for $\theta$, which is $(\hat\theta_B^{\ast(\alpha)},\,\hat
\theta_B^{\ast(1-\alpha)})$. According
to Efron \& Tibshirani (1993), the number of independent repetitions 
of the bootstrap sampling shall be set as $B=2000$.

\section{Mass-Period Correlations}

We first work on the mass-period correlation in $M-P$ space. With
the 233 observed data, we calculate the correlation coefficients
and use the bootstrap method to determine the confidence
intervals. We find that the correlation coefficient between $M$
and $P$ is 0.1762 and the resulting 95\% bootstrap confidence
interval is $(0.0575, 0.3130)$. This indicates that the mass $M$
and period $P$ has a weak positive correlation.
Fig. 5 shows the 233 samples in $M-P$ space. In deed, it looks
as that the distribution is not completely random. However, 
the positive correlation is difficult to recognize and so it is 
consistent with a weak correlation.

On the other hand, for the mass and period distributions in
 $\ln M$-$\ln P$ space, the correlation coefficient between $\ln M$
and $\ln P$ is $0.3876$ and the corresponding 95\% bootstrap
confidence interval is $(0.2668, 0.5001)$. It clearly indicates
that there is a positive mass-period correlation for the data in
$\ln M$-$\ln P$ space.



Fig. 6 shows the 233 exoplanets in $\ln M$-$\ln P$ space
and it seems there are many fine structures of data distributions.
For example, for those points with $\ln M > 0$,  it is far more crowded
in the region with $\ln P > 5$. This is, of course, related to
the deficit of massive close-in planets, which makes the main contribution
on the positive mass-period correlation.
On the other hand, partially due to the effort of transit surveys,
the discovery of many hot Jupiters also makes it a bit crowded around 
$(\ln M, \ln P) = (-0.5, 1)$. However, 
the deficit of sub-Jupiter mass planets at separations about
0.5 AU mentioned in Papaloizou \& Terquem (2006) seems to disappear
in this plot of 233 data points.

In fact, the mass-period correlation was 
theoretically studied in P\"atzold \& Rauer (2002) and Jiang et
al.(2003), which focused on the paucity of massive close-in planets.
The explanation for this paucity  and the correlation could be that
the tidal interactions with host stars make the massive close-in
planets migrate inward and finally merge with the stars.

\section{Concluding Remarks}

In this paper, we first construct a new
mass-period distribution function
of exoplanets, in which the correlation is considered.
This is the first time in this field and was not possible until
the method was proposed in Magnussen (2004).

The correlation coefficients of exoplanet data in  $M$-$P$ space
and  $\ln M$-$\ln P$ space are further determined
and the
bootstrap method is then used to construct the confidence
intervals of correlation coefficients at particular confidence
levels.
We confirm that there is a mass-period
correlation for exoplanets.
In addition to the paucity of massive close-in planets, there are
other fine structures in the data distribution to be
investigated in the future.

\clearpage
\begin{figure}
\epsscale{1.0}
 \plotone{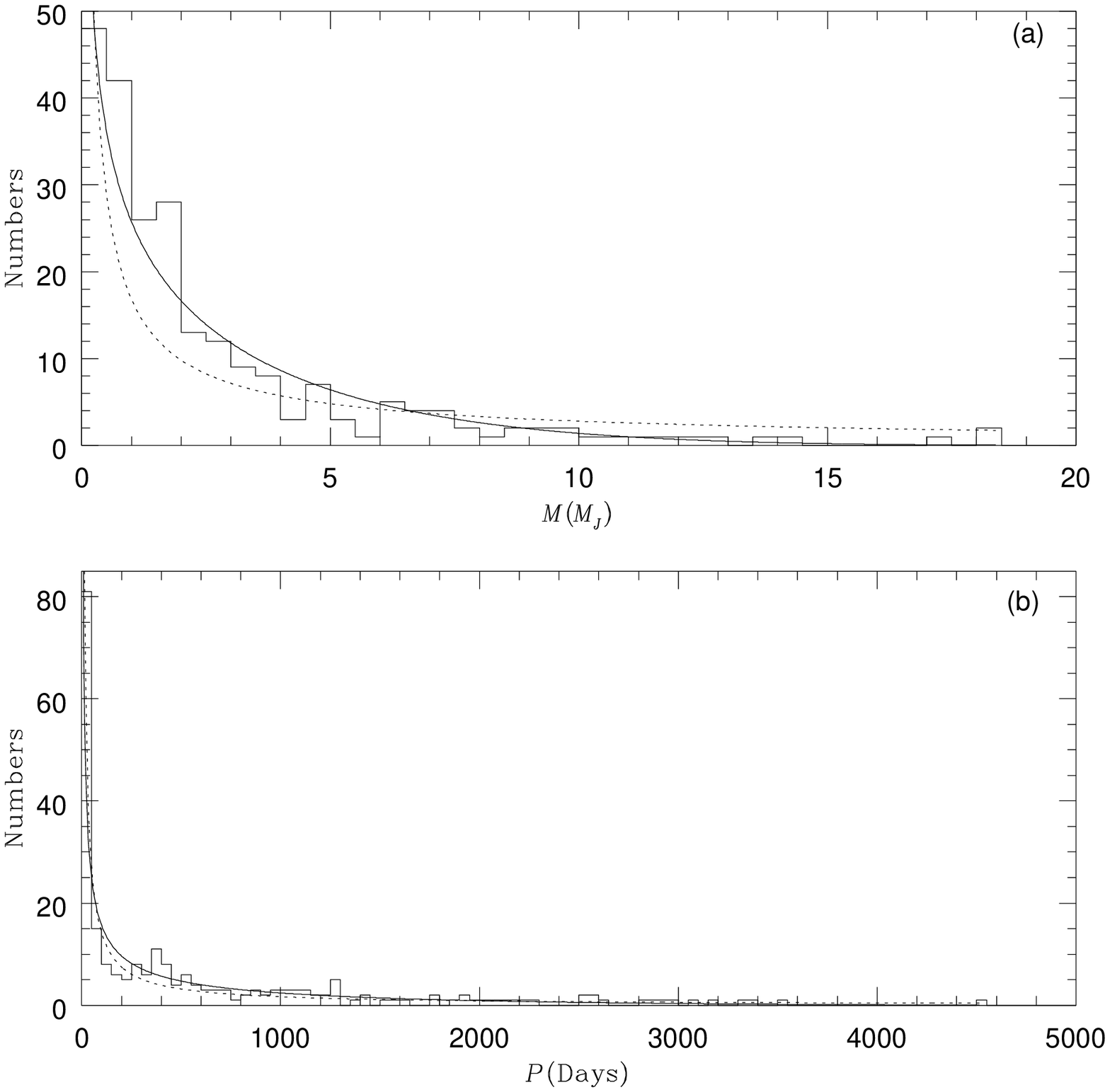}
  \caption{The mass and period histograms:
(a) the mass histogram of 233 exoplanets, where the solid curve is
$f^{his_M}_{beta}$ and the dotted curve is
$f^{his_M}_{power}$; (b) the period histogram of 233 exoplanets,
where the solid curve is $f^{his_P}_{beta}$ and the dotted curve
is $f^{his_P}_{power}$.  Please note that the mass's unit is 
Jupiter Mass ($M_J$) and the period's unit is days. 
The bin size of 
the above mass histogram is 0.5 $M_J$ and the bin size of the
above period histogram is 50 days. 
}
\end{figure}

\clearpage
\begin{figure}
\epsscale{1.0}
 \plotone{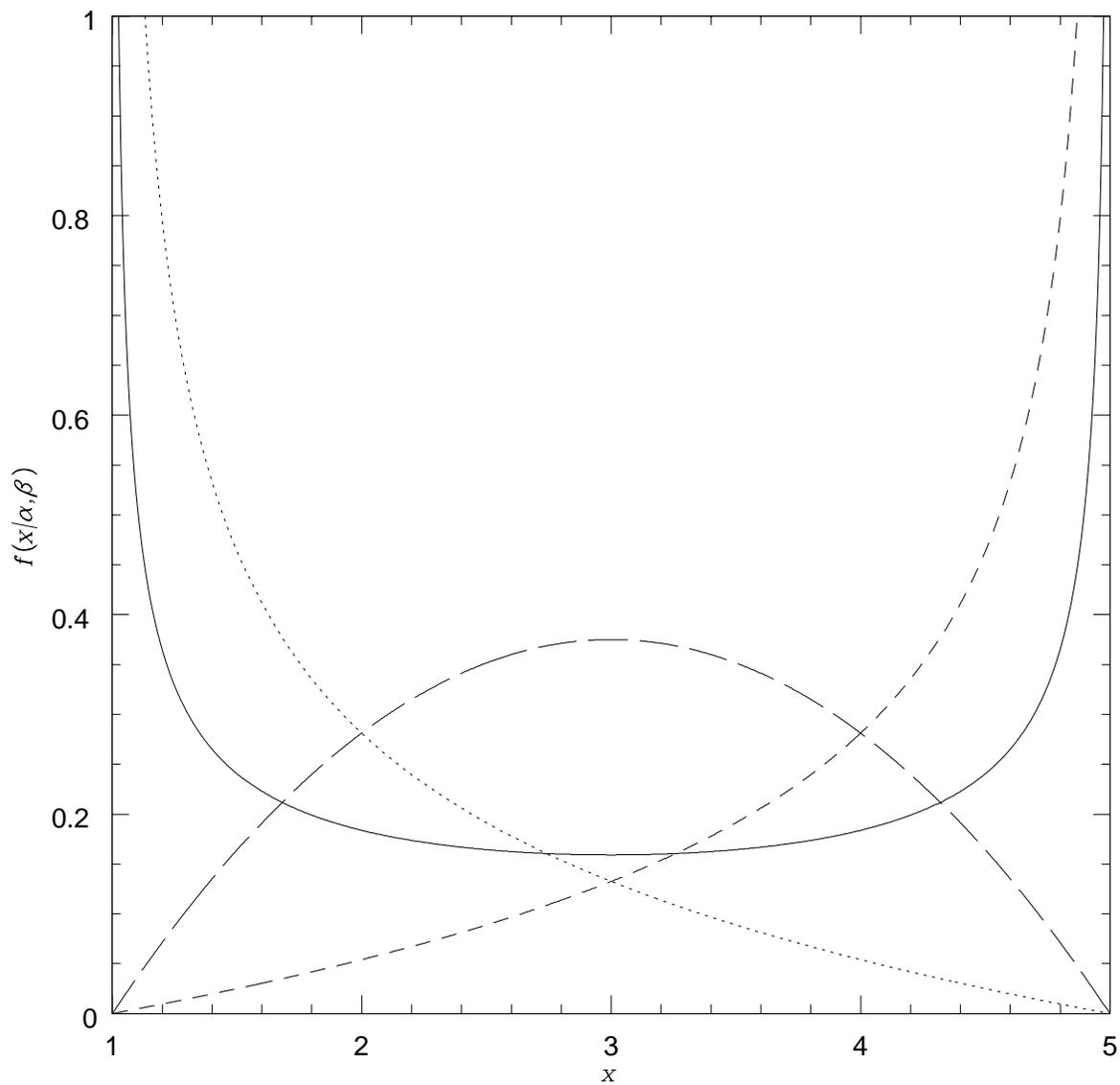}
 \caption{Beta
distribution functions in Eq.(3) with $c=1$, $d=5$: the solid
curve is for $\alpha=\beta=0.5$; the dotted curve is for
$\alpha=0.5, \beta=2$; the short dashed curve is for $\alpha=2,
\beta=0.5$; and the long dashed curve is for $\alpha=\beta=2$.}
\end{figure}

\clearpage
\begin{figure}
\epsscale{1.0}
\plotone{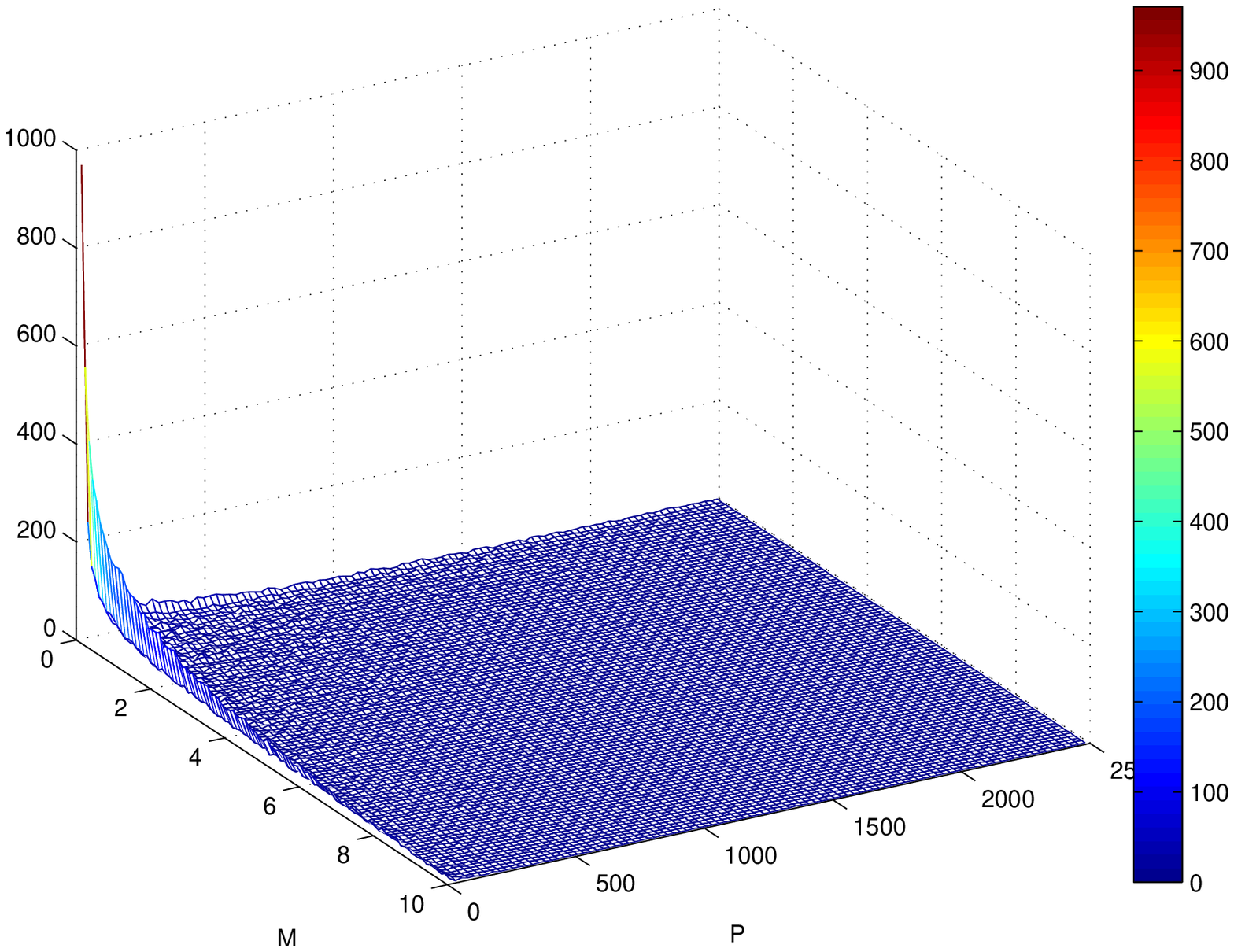}
\caption{The correlated mass-period distribution function. 
The unit of the mass $M$ is 
Jupiter Mass ($M_J$) and the unit of the period $P$ is days. 
}
\end{figure}

\begin{figure}
\epsscale{1.0}
 \plotone{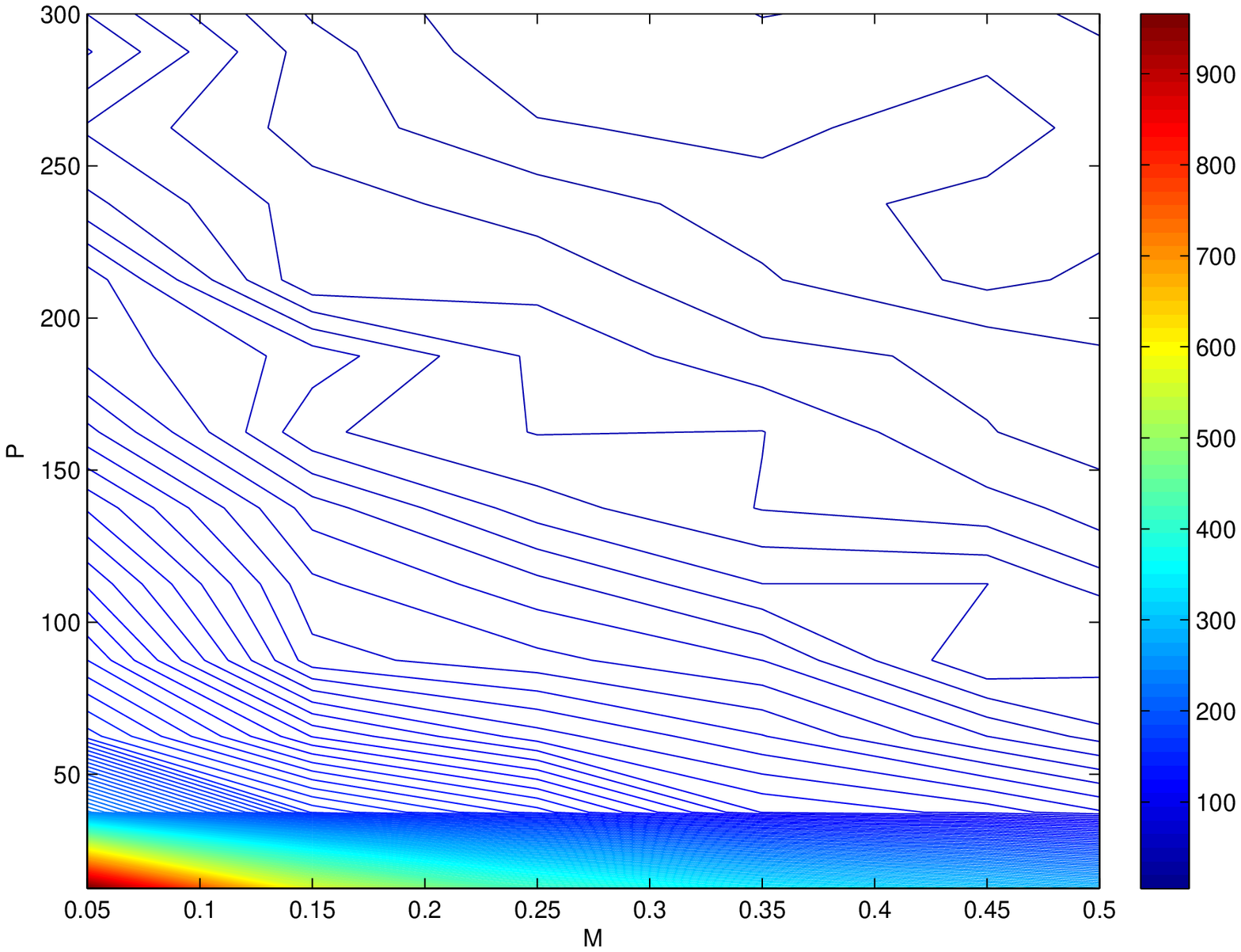}
  \caption{The contour of the correlated mass-period distribution function.
The unit of the mass $M$ is 
Jupiter Mass ($M_J$) and the unit of the period $P$ is days. 
}
\end{figure}

\begin{figure}
\epsscale{1.0}
 \plotone{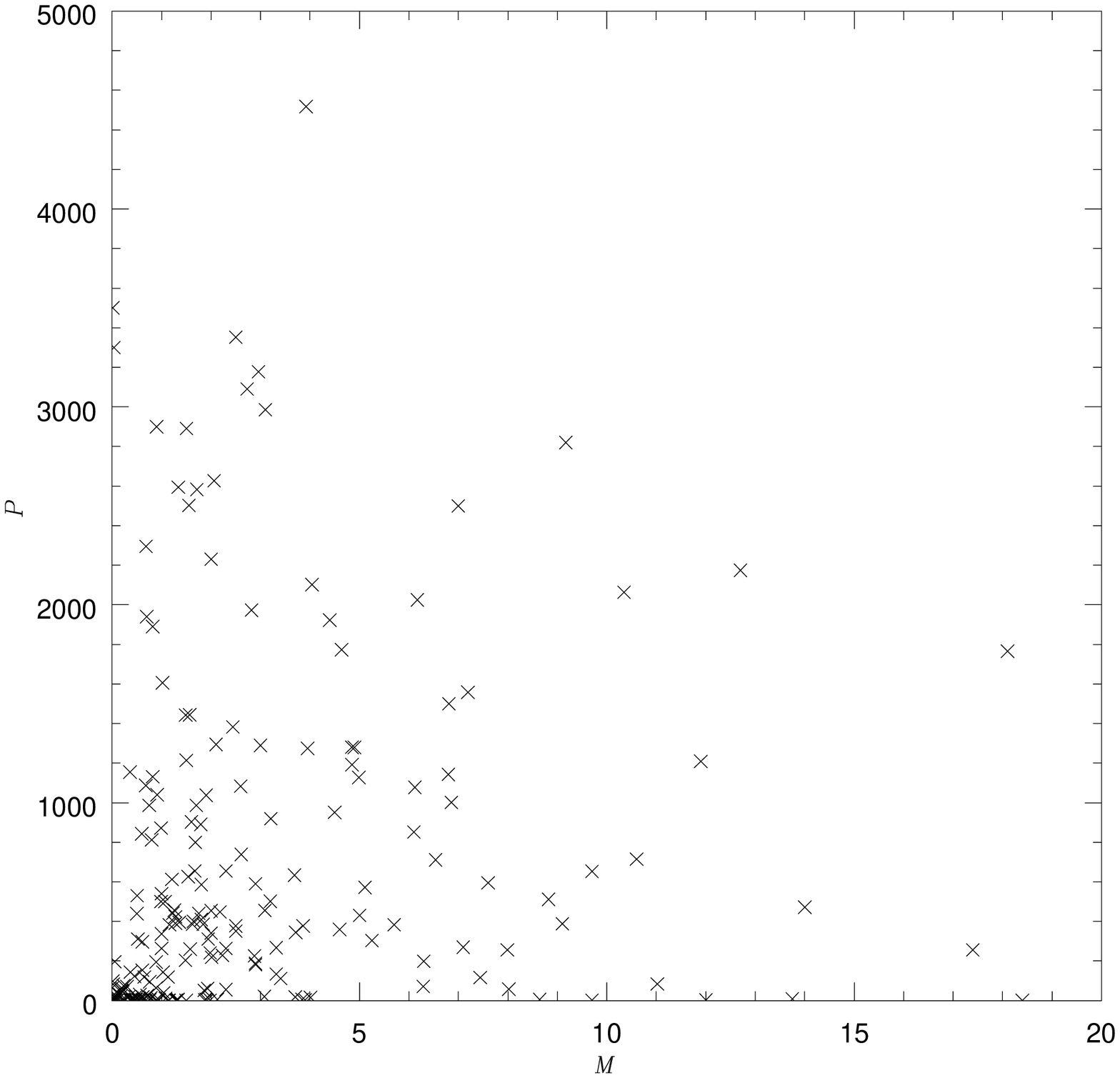}
  \caption{The data of 233 exoplanets in $M-P$ space.
The unit of the mass $M$ is 
Jupiter Mass ($M_J$) and the unit of the period $P$ is days. 
}
\end{figure}

\begin{figure}
\epsscale{1.0}
 \plotone{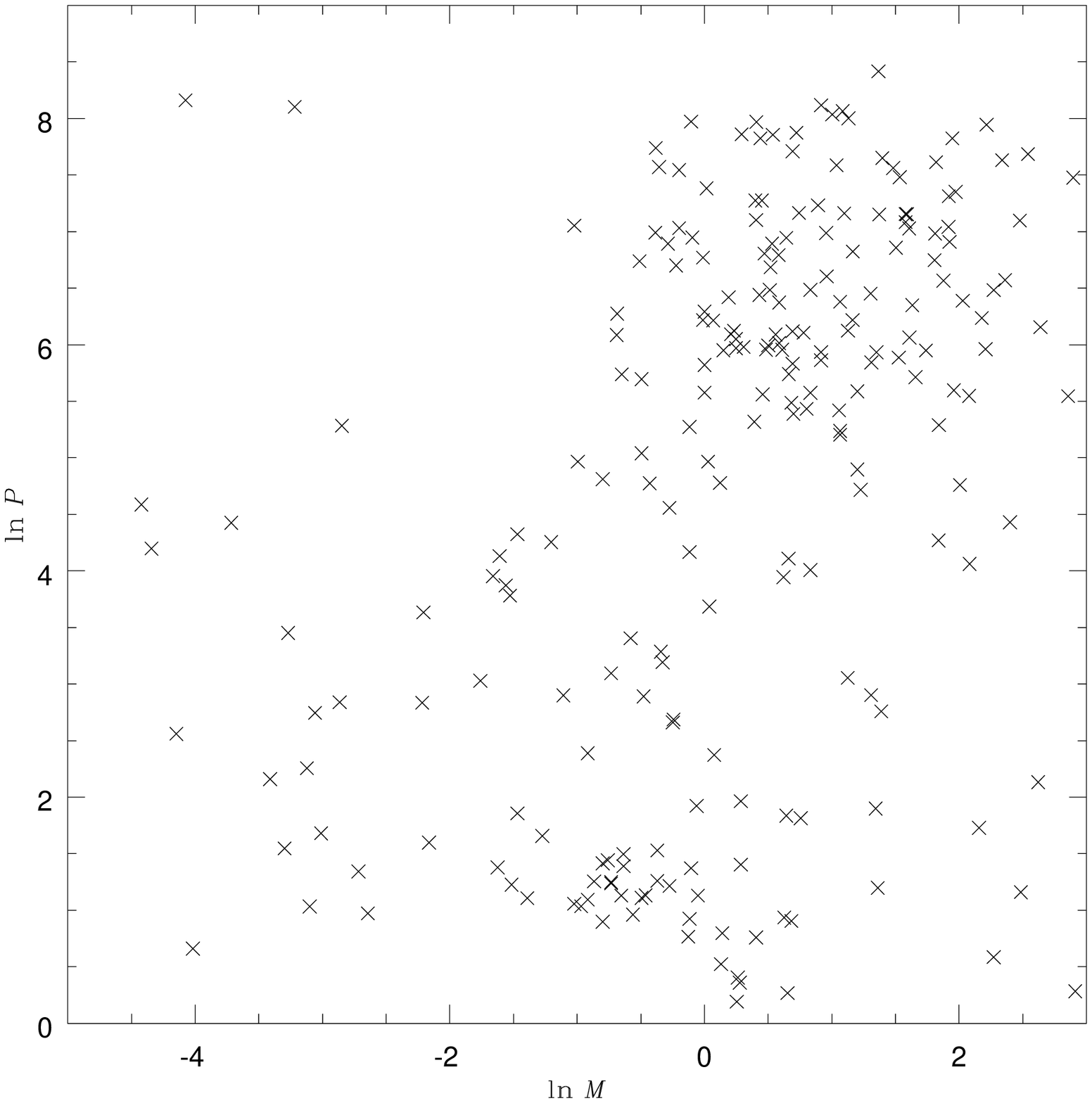}
  \caption{The data of 233 exoplanets in ${\rm ln}M - {\rm ln}P$ space.
The unit of the mass $M$ is 
Jupiter Mass ($M_J$) and the unit of the period $P$ is days. 
}
\end{figure}

\end{document}